\documentclass[pre,showkeys,twocolumn,showpacs,amsmath,amssymb]{revtex4-1}
\usepackage[T1]{fontenc}
\usepackage{color}
\usepackage[latin1]{inputenc}
\usepackage{graphicx}
\usepackage{bbold}
\usepackage{epsfig}
\usepackage{rotating}
\usepackage{dcolumn}
\usepackage{bm}
\usepackage{tikz}
\usepackage{amsmath}
\usepackage{amssymb}

\newcommand{\tikzcircle}[2][red,fill=red]{\tikz[baseline=-0.5ex]\draw[#1,radius=#2] (0,0.02) circle ;}%
\newcommand{\tikzsquare}[2][red,fill=red]{\tikz[baseline=-0.5ex]\draw[#1,radius=#2] (0.25,0.15) rectangle (0,-0.1);}
\definecolor{myblack}{rgb}{0.0, 0.0, 0.0}
\definecolor{myred}{rgb}{1.0, 0.0, 0.0}
\definecolor{mygray}{gray}{0.6}

\begin{document}

\title{Clustering in vibrated monolayers of granular rods}

\author{M. Gonz\'alez-Pinto}
\email{miguel.gonzalezp@uam.es}
\affiliation{Departamento de F\'{\i}sica Te\'orica de la Materia Condensada,
Universidad Aut\'onoma de Madrid,
E-28049, Madrid, Spain}

\author{F. Borondo}
\email{f.borondo@uam.es}
\affiliation{Departamento de Qu\'{\i}mica, Universidad Aut\'onoma de Madrid, E-28049, Madrid, Spain\\
Instituto de Ciencias Matem\'aticas (ICMAT),
Universidad Aut\'onoma de Madrid, E-28049, Madrid, Spain}

\author{Y. Mart\'{\i}nez-Rat\'on}
\email{yuri@math.uc3m.es}
\affiliation{Grupo Interdisciplinar de Sistemas Complejos (GISC), Departamento de Matem\'aticas, Escuela
Polit\'ecnica Superior, Universidad Carlos III de Madrid, Avenida de la Universidad 30, E-28911, Legan\'es, Madrid, Spain}

\author{E. Velasco}
\email{enrique.velasco@uam.es}
\affiliation{Departamento de F\'{\i}sica Te\'orica de la Materia Condensada,
Instituto de F\'{\i}sica de la Materia Condensada (IFIMAC) and Instituto de Ciencia de Materiales Nicol\'as Cabrera,
Universidad Aut\'onoma de Madrid,
E-28049, Madrid, Spain}

\date{\today}

\begin{abstract}
We investigate the ordering properties of vertically-vibrated monolayers of granular cylinders in a 
circular container at high packing fraction. In line with previous works by other groups, we identify 
liquid-crystalline ordering behaviour similar to that of two-dimensional hard rectangular particles 
subject to thermal equilibrium fluctuations. However, due to dissipation, there 
is a much stronger tendency for particles to cluster into parallel arrangements in the granular system.  
These clusters behave as a polydisperse mixture of long life-time `superparticles', and some aspects 
of the system behaviour can be understood by applying mean-field theories for equilibrium hard rectangles, 
based on two-body correlations, to these `superparticles'.  Many other features of the granular system 
are different: (i) For small particle length-to-breadth ratio $\kappa$, we identify tetratic ordering at 
moderate packing fractions and smectic fluctuations at higher packing fractions, with no sharp transition 
between the two states. Both types of ordering can be explained in terms of clustering.
(ii) For large $\kappa$, strong clustering precludes the stabilisation of a uniaxial nematic state, and the system
exhibits a mixture of randomly-oriented clusters which, as packing fraction is increased, develops into
states with smectic fluctuations, again through 
a diffuse transition. (iii) Vorticity excitations of the velocity field compete with smectic ordering, 
causing dynamic fluctuations and the absence of steady states at high densities; the tetratic state, 
by contrast, is very stiff against vorticity, and long-standing steady states, spatially and orientationally homogeneous except for four symmetrical defects located close to the wall, can be observed.
\end{abstract}


\maketitle

\section{Introduction}

Thermal systems of particles driven by exclusion forces are governed solely by entropy and show interesting 
behaviours \cite{review}. Hard spheres and discs crystallise at high packing fraction $\varphi$, while 
dense systems of anisometric particles can also form oriented fluid phases (nematics) when the particle 
aspect (length-to-breadth) ratio $\kappa$ is larger than a critical value \cite{Frenkel}. In the uniaxial nematic phase the 
long axes of the particles tend to orient parallel to each other and a globally oriented direction (the director) 
results. When $\kappa$ is low, two-dimensional particles of rectangular shape form instead a different 
nematic phase, the {\it tetratic} phase \cite{schlacken}, an exotic biaxial phase with two (instead of one) 
perpendicular, equivalent directors. The stability of the tetratic phase stems from the particularly favourable perpendicular 
arrangements of particles with sharp corners. For high $\varphi$ particles form phases with partial 
(and eventually full) spatial order, although the exact phase diagram of, in particular, hard rectangular 
particles, and the dependence with aspect ratio, has not been studied in detail.

Horizontally vibrated monolayers of granular, dissipative particles have been shown to exhibit patterns that 
resemble those of thermal equilibrium systems \cite{hindues2,hindues1,hindues,Aranson,galanis1,galanis2,daniel}. 
Granular rods of various shapes show liquid-crystal 
arrangements at high $\varphi$. In one recent experiment \cite{daniel}, the transition from a fully 
disordered (isotropic) phase to the uniaxial nematic or tetratic phase has been characterised from the 
behaviour of standard order parameters obtained from adequate averages of particle orientation distributions. 
Depending on the aspect ratio, the isotropic was shown to order into a tetratic (low $\kappa$) or 
uniaxial nematic (high $\kappa$) phase as $\varphi$ was increased. The critical aspect ratio $\kappa^*$
that separates the two ordered phases was estimated to be $\kappa^*\simeq 7.3$. This is much higher than 
the value predicted by mean-field theories ($\kappa^*=2.2$ or $3.2$, depending on whether three-particle
correlations are neglected or not \cite{SPT,3body}), but close to the results from equilibrium computer simulations 
\cite{daniel}. However, the value of packing fraction at the isotropic-uniaxial nematic transition in the granular
system shows qualitatively significant deviations with respect to the equilibrium 
(either mean theory or simulation) calculations, indicating the existence of genuine non-equilibrium effects 
associated with dissipation. Clearly, in this system entropic or exclusion interactions are strongly 
coupled to dissipation induced by friction and inelastic collisions, and it remains to be seen to what extent 
entropy-driven and packing effects compete with hydrodynamic granular flow and dissipation to produce the observed 
ordered patterns. At least for hard spheres, there seem to exist some regimes where mean-field theories and computer 
simulations for equilibrium thermal hard particles (i.e. methods based on entropy maximisation) are useful in discussing 
results from granular systems. For example, radial distribution functions of dense systems of granular spheres 
and discs are quantitatively similar to the corresponding equilibrium systems \cite{gdr}, but granular discs
show a more complex phase behaviour and a strong phase dependence with respect to the shaking frequency and 
amplitude \cite{Urbach}. Phase ordering of granular anisometric particles may also be strongly affected by the 
manner in which energy is injected into the system and dissipated.

In this paper we report on experiments using vibrated monolayer of granular cylindrical particles, which project 
on the horizontal plate approximately as hard rectangles. Our aim is to assess to what extent the ordered liquid-crystalline 
patterns predicted for thermal hard rectangles in equilibrium can be observed in the granular experiments. In doing 
so, we have found a strong dependence on particle aspect ratio of the liquid-crystalline patterns observed, as 
previously obtained \cite{daniel}. We have found that the tendency for clustering of particles is 
very strong, and that these effects help explain most of the phenomenology observed, in particular, some of the differences 
with respect to equilibrium systems. Thus far, clusters and 
clustering effects in general have not been carefully analysed in experiments on dense packings of vibrated
granular rods. We argue that clustering gives
a useful picture to explain the different high-density configurations with respect to aspect ratio.
Dense collections of dissipative rectangular particles exhibit a strong tendency to lie parallel 
to each other, and clusters of approximately parallel particles are formed easily. These clusters are very stable, and 
in many respect behave as single units, which we call `superparticles'. The clustering
tendency is much stronger than in fluids of equilibrium thermal rods, as demonstrated by comparison with
equilibrium Monte Carlo simulations on thermal particles. In the vibration experiments, 
particles of low aspect ratio, $\kappa<8$, are seen to form clusters with square or close-to-square shape quite easily, 
and these superparticles of still lower effective aspect ratio may orient parallel or perpendicular in relatively 
efficient packing configurations; bulk tetratic ordering, and smectic at even higher packing fractions, may result. 
However, for long rods, $\kappa>8$, large clusters involve many particles and tend to bend, so that superparticles cannot
pack into tetratic or uniaxial nematic configurations: randomly oriented arrangements of large clusters 
or `patches' are preferred, which at high density also exhibit smectic fluctuations. 
In addition, this granular system features the usual non-equilibrium properties of vibrated monolayers: strong 
dependence with excitation parameters, formation of cavities and large density fluctuations, hydrodynamic flow 
competing with ordering, etc. In particular, we do not observe sharp transitions between states with different 
ordering properties, a result which is at variance with those of Ref. \cite{daniel}. However, the experiments
may not necessarily be in contradiction,
given that the experimental conditions of both experiments are not identical and, in addition, different protocols to
calculate the order parameters could have been used. Another distinct feature of our experiment is that, in the high-density 
regime, there is an absence of well-defined steady states, which we 
explain as a competition between smectic ordering effects due to packing and disordering effects due to excitation of 
hydrodynamic modes. Finally, an intriguing result is that, as observed in Ref. \cite{daniel}, $\kappa^*=7$--$8$ is a critical aspect
ratio separating two types of orientational ordering, similar to equilibrium. But clustering effects and the associated
formation of superparticles renormalise this value to $\kappa^*\simeq 2.1$, a value very close to that predicted by
equilibrium mean-field theories. We suggest that clustering renormalises the range of interactions, which are reduced
to the mean-field, two-particle level.

In Section \ref{exp} details on the experiment and the parameters used are given.
Section \ref{results} presents the results from the cluster analysis, the order parameters and the angular
distribution functions, both for rectangles and for superparticles. Section \ref{discussion} presents a
discussion and our interpretation of the results, and also contains some concluding remarks.
An explanation on the analysis tools are relegated to the Appendix.

\section{Experiment and analysis}
\label{exp}

Our experiment uses nonmagnetic steel cylinders of length $L$ and breadth $D=1$ mm. Plastic 
cylinders were also studied in one case for comparison \cite{plastic}. The packing fraction is calculated as
$\varphi=NLD/A$, where $N$ is the number of cylinders and $A$ the area of the circular cavity.
The vibration experiment is standard:
particles are inside a cylindrical container, consisting of a thin horizontal cylindrical cavity made of aluminium,
with a diameter of 14 cm. An upper plastic lid limits the free motion of cylinders in the vertical direction, 
which is $\simeq 0.8D$, so that particles cannot overlap in the plane of the plate. 
Vertical vibration is induced by an electromagnetic shaker which generates a 
sine-wave vertical motion of frequency $\nu$ and amplitude $a$. The value of frequency is set to $\nu=39$ Hz.
Effective accelerations $\Gamma=a\nu^2/g$ in the range $3 - 4$ are used. Results are observed to be rather sensitive to the frequency, 
but so much to the value of $\Gamma$. Some frequencies 
tend to promote the formation of cavities, i.e. persistent regions with very few particles \cite{hindues}, even in the 
high-packing-fraction regime. We have tried to avoid these states, as well as the dilute regime, by always choosing 
$\varphi\agt 0.50$.

\begin{figure}
\begin{center}
\includegraphics[width=0.90\linewidth,angle=0]{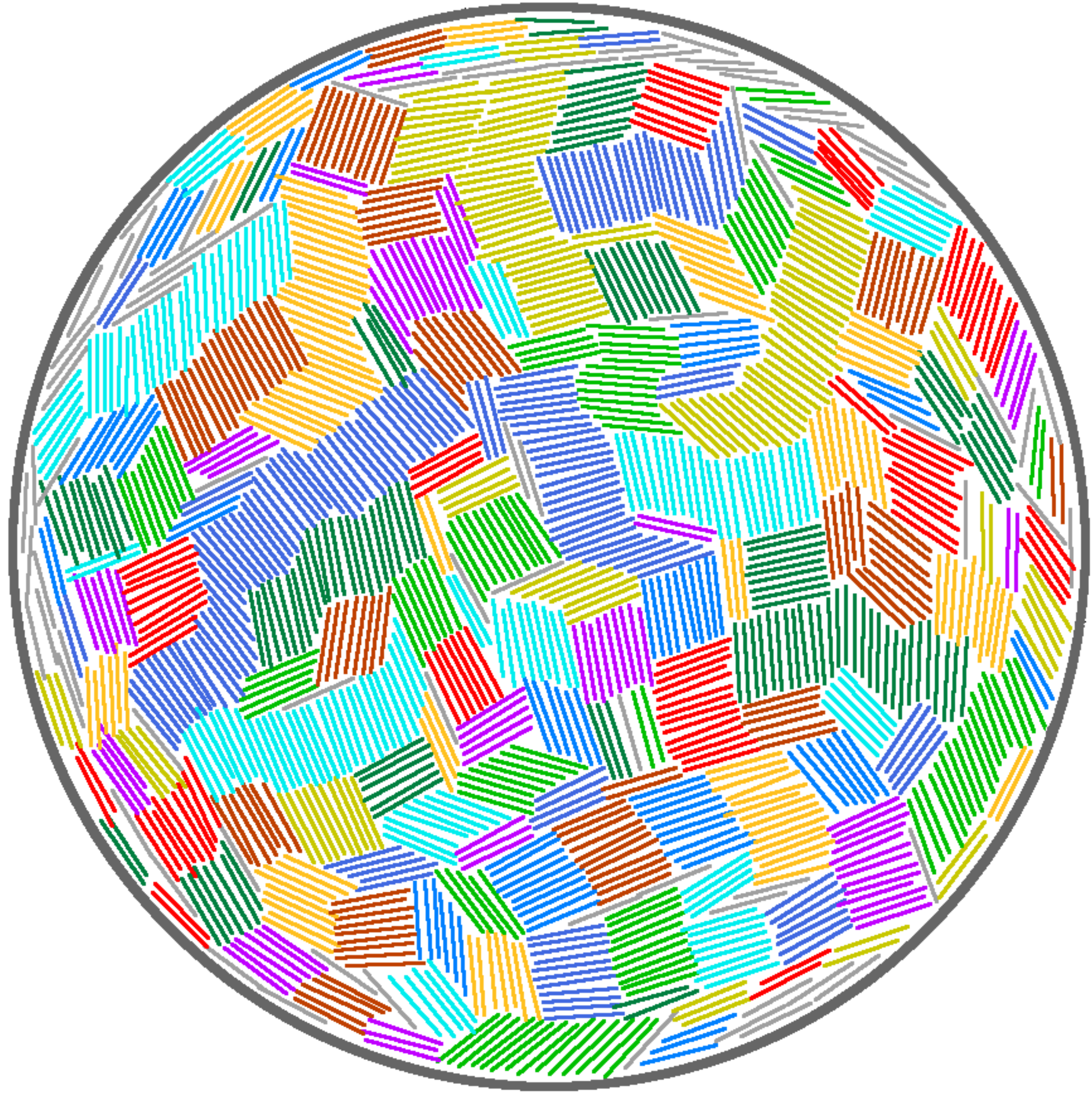}
\caption{Snapshot of particle configurations of a system with $\kappa=10$ at packing fraction $\varphi=0.82$. 
Particles belonging to the same cluster are drawn as straight lines of the same colour.}
\label{fig1}
\end{center}
\end{figure}

To analyse the experiments we use standard procedures. All measured quantities are obtained from the
projected coordinates of the cylinders on the plate, ${\bm r}=(x,y)$,
and from the projected orientations of the long particle axis 
$\hat{\bm e}$ of each particle. These are obtained by using the ImageJ software \cite{imageJ}. Typically 98\%
of particles are successfully identified. Particle velocities 
${\bm v}$ are calculated by identifying particle positions in two consecutive frames, separated by 0.3 s, and using a 
simple finite-difference approximation. Success in the identification of particle trajectories decreases from 98\% for 
$\kappa\le 8$ to 85\% (at worst) for $\kappa=4$; the latter case, however, involves a larger number of particles, so that
statistical errors are not significantly altered. Calculation of velocities allows for the computation of a local 
granular temperature (see Sec. \ref{results}). The orientational and positional orders in the cell are quantified by the local 
and global uniaxial nematic, $q_2$,
tetratic, $q_4$, and smectic, $\sigma_S$, order parameters. The nematic and tetratic order parameters were obtained 
through the orientational distribution $h(\phi)$, and also through properly defined order tensors. Details on the definition
and calculations of these parameters are given in the Appendix. 

Finally, in order to study collective properties of the cylinders from a different perspective,
we analysed clustering properties. This has not been done systematically in vibrated experiments on rods,
but we have concluded that these properties are extremely useful. We define a cluster as a collection of
cylinders that are approximately parallel and spatially close to each other. Our connectedness criterion
imposes two conditions for two cylinders to be connected: (i) their projected long axes form an angle
$\Delta\phi=\phi-\phi' < 10^{\circ}$, and (ii) their projected geometric centres are at a distance $d< 1.8 D$
(we should mention that, as usual in cluster analysis, results turn out not to depend qualitatively on the exact 
definition of the connectedness criterion). Once the connection matrix for the cylinders is established, a standard 
routine to identify the cluster distribution $N_c(n)$ is used to obtain the average number of clusters $N_c$ of 
a given size $n$, where $n$ is the number of cylinders making up the cluster. 

\section{Results}
\label{results}

In this section we present results for the different quantities defined in the previous section.
In all the experiments particles were initially placed in the container
by hand, trying to achieve a random initial arrangement.

\subsection{Clustering}

We start by looking at the particle distribution from the point of view of cluster formation. 
Clustering effects play an important role in dense packings of dissipative rectangular particles. 
Clustering is known to enhance the tendency for tetratic 
ordering in dense packings of particles with low $\kappa$ at thermal equilibrium. 
Monte Carlo simulations on equilibrium rectangular rods demonstrated this clustering 
effect \cite{Granada}. A theory based on stable, polydisperse clusters with a distribution decaying exponentially with 
size was shown to lead to good agreement with simulations of equally-sized rectangles \cite{PRE}. It is therefore worthwhile to discuss the 
cluster statistics of our experiments. 

The formation of clusters is quite apparent in our dissipative vibrated system. Fig. \ref{fig1} shows a picture 
of a typical configuration of a system with $\kappa=10$ and $\varphi=0.82$. Particles belonging to the same cluster, 
as defined by the connectedness criterion mentioned in Section \ref{exp}, are drawn in the same colour. Well-packed 
clusters are formed. The dynamics of a given cluster can be followed for a significant fraction of the total 
experimental time.

\begin{figure}
\begin{center}
\includegraphics[width=0.95\linewidth,angle=0]{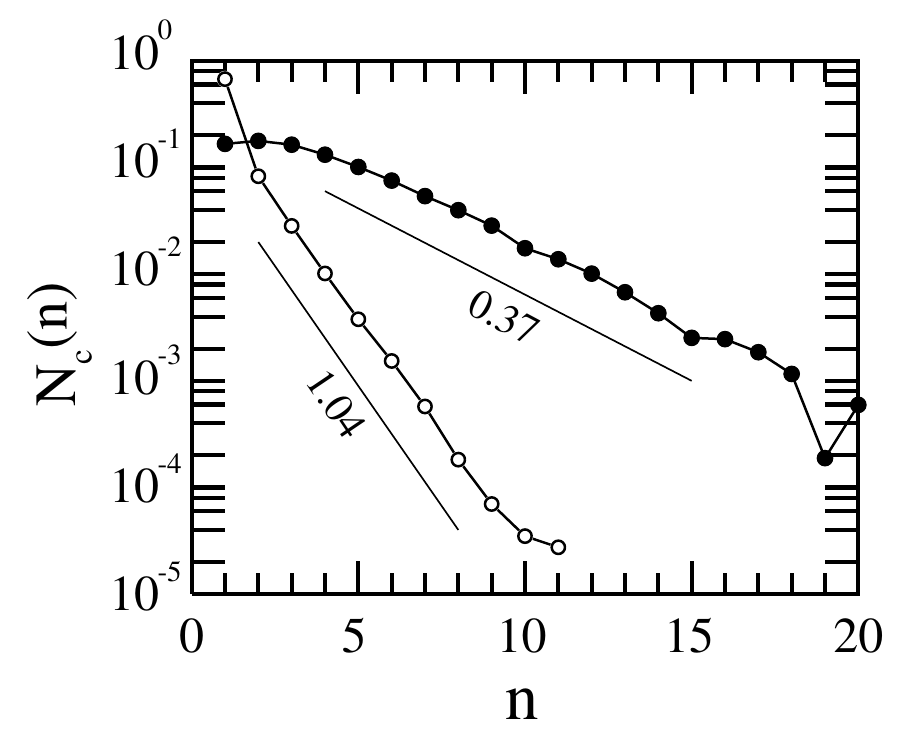}
\caption{Normalised cluster distribution functions $N_c(n)$ as a function of cluster size $n$ for $\simeq 2000$ 
particles with
$\kappa=4$ at packing fraction $\varphi=0.52$ in a circular container of diameter $14$ cm. Open circles: Monte
Carlo simulation at thermal equilibrium. Filled circles: vibration experiment. Straight lines represent reference exponential functions with values for
decay parameters as indicated.}
\label{clustering}
\end{center}
\end{figure}

Fig. \ref{clustering} shows the cluster distribution function $N_c(n)$ from a real vibration experiment compared to an equivalent Monte Carlo simulation of hard rectangles in a circular cavity. The aspect ratio in both cases 
is $\kappa=4$, the packing fraction is set to $\phi=0.52$ and the number of particles was adjusted to be equal in both experiment
and simulation. Although both distributions are close to exponential, $N_c(n)\propto e^{-\lambda n}$, we can see that the cluster
distribution of the granular
system decays more slowly with cluster size (approximately a factor 3 in decay parameter), indicating the stronger tendency 
of the granular system to cluster than in the thermal equilibrium system. 

\begin{figure}
\begin{center}
\includegraphics[width=1.00\linewidth,angle=0]{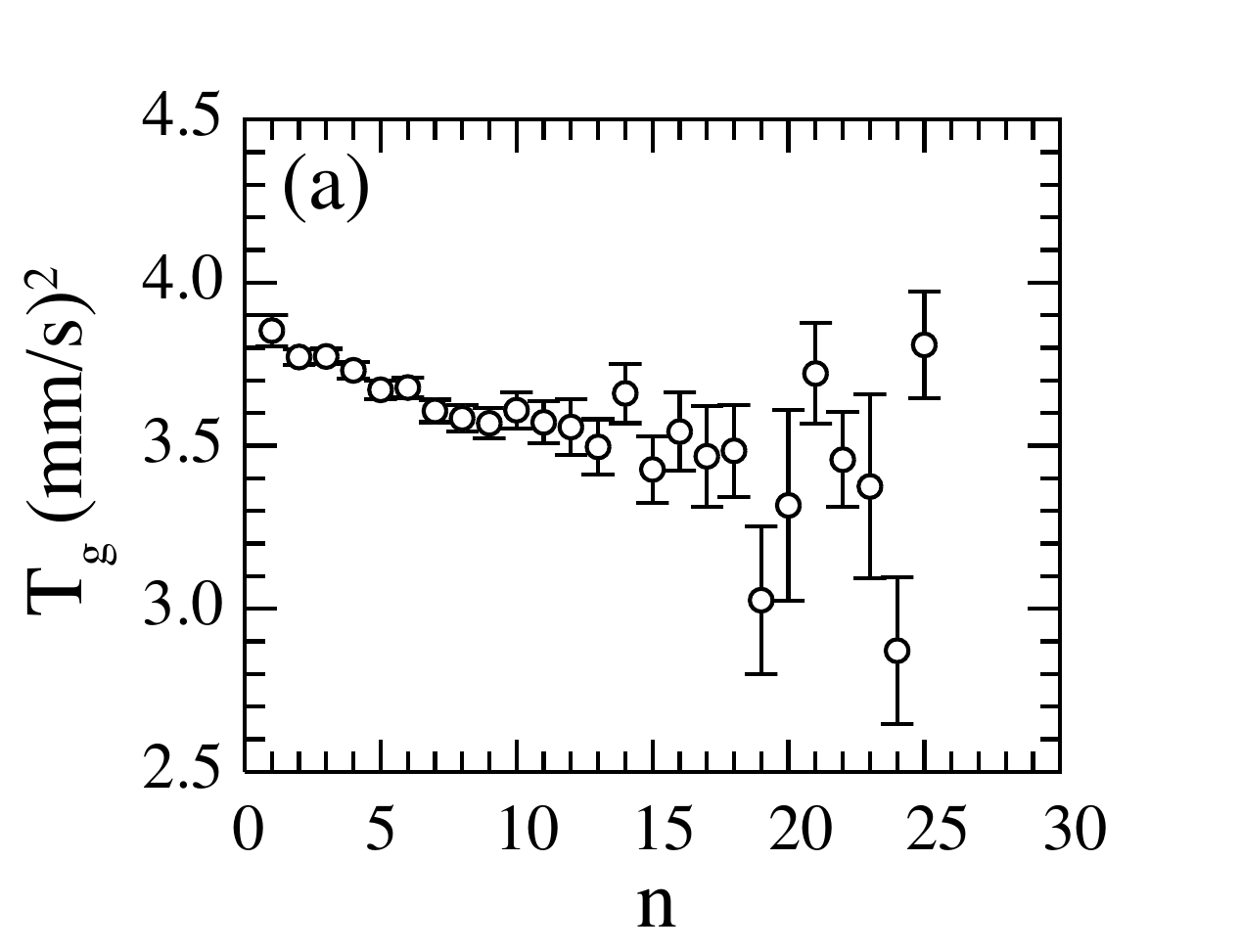}
\includegraphics[width=1.00\linewidth,angle=0]{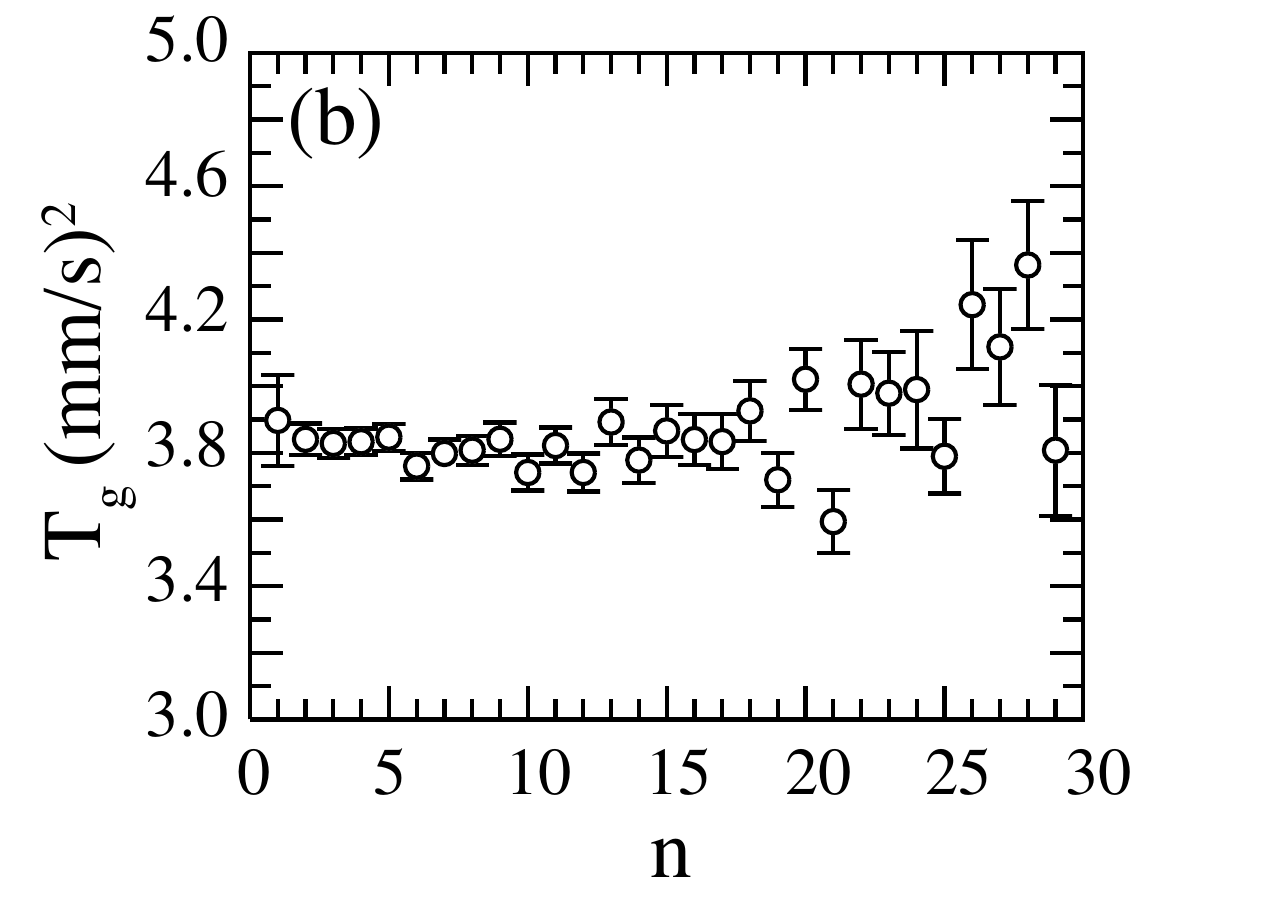}
\caption{Cluster granular temperature $T_g$ as a function of cluster size for two different conditions. 
(a) $k=8$, $\varphi=0.78$. (b) $k=10$, $\varphi=0.70$. Error bars indicate the mean-square fluctuation
of each data point.}
\label{new}
\end{center}
\end{figure}

Clustering of dissipative particles stems 
from the locally cooling effect of inelastic particle collisions and the associated tendency of particles to align in 
highly dense packing configurations. Contrary to the equilibrium system, dissipation in the granular fluid can lead
to local inhomogeneities of the granular temperature.
Since particles organise into long-lived clusters, one could think that local inhomogeneities caused by the local cooling 
effect could give different granular temperatures for clusters of different sizes. We define the granular 
temperature $T_g(n)$ for clusters of size $n$ from the mean square velocity of the particles in the cluster, 
\begin{eqnarray}
T_g(n)=\left<\frac{1}{n}\sum_{i=1}^n\left|{\bm v}_i-{\bm V}\right|^2\right>_{n,T},
\end{eqnarray}
where ${\bm v}_i$ is the velocity of a particle in the cluster, ${\bm V}$ the local velocity of the cluster centre of mass,
and $\left<\cdots\right>_{n,T}$ is an average over clusters of size $n$ and over time. Fig. \ref{new} shows the
cluster granular temperature as a function of cluster size for two different conditions. Although we have observed that
in $\sim 80\%$ of the experiments $T_g(n)$ decreases with $n$, as in panel (a), the effect is not robust, since in the 
remaining cases $T_g(n)$ hardly changes with size, as in panel (b).

\begin{figure*}
\begin{center}
\includegraphics[width=1.0\linewidth,angle=0]{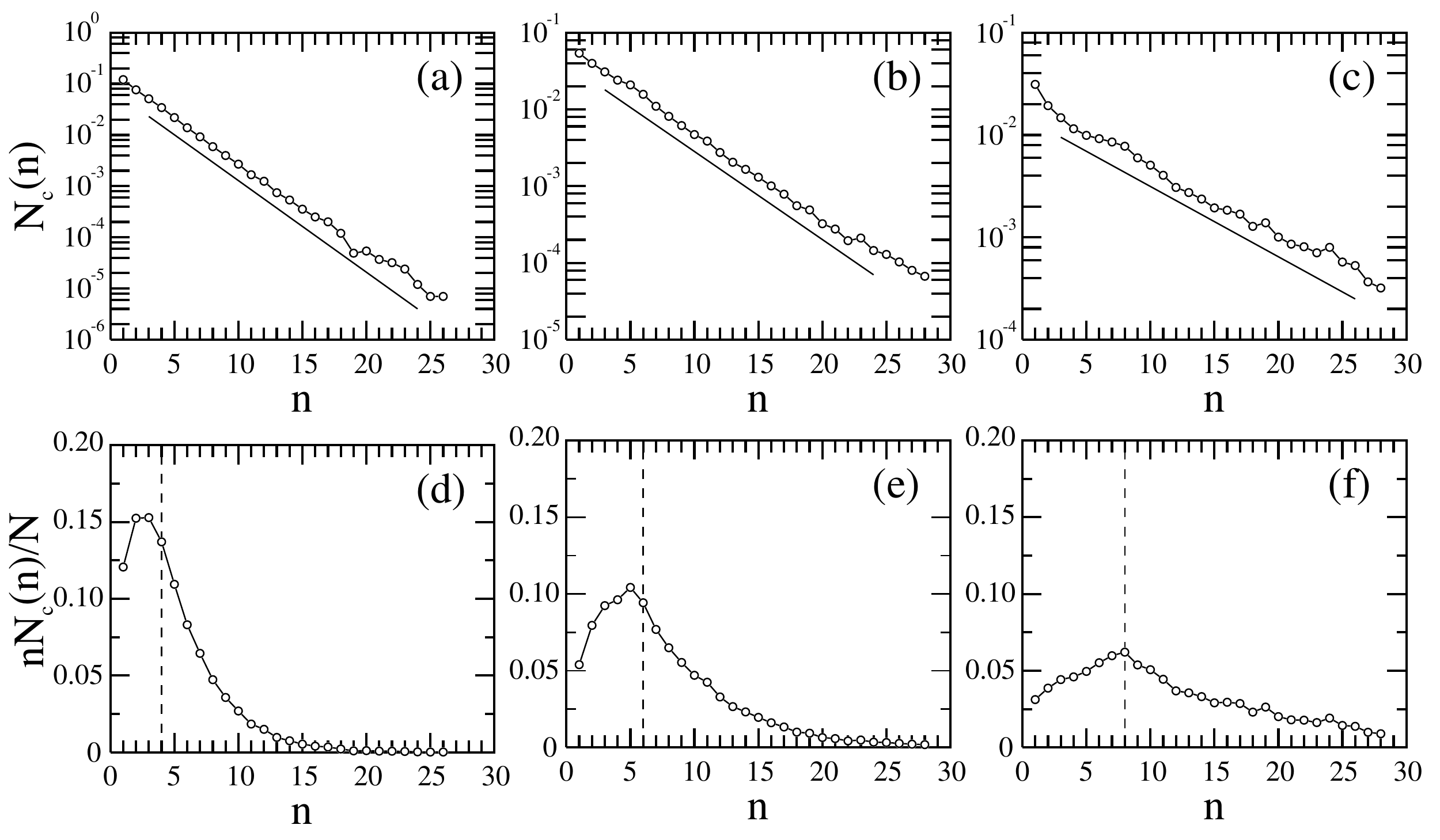}
\caption{Plots in (a-c) give the normalised cluster distribution functions $N_c(n)$ as a function of cluster size $n$ for 
(a) $\kappa=4$ and $\varphi=0.59$; (b) $\kappa=6$ and $\varphi=0.69$; and (c) $\kappa=8$ and $\varphi=0.78$.
Straight lines correspond to expotential behaviour (for reference). Plots in (d-f) are the fraction
of particles in clusters of size $n$. (d) conditions as in (a); (e) conditions as in (b); and (f) conditions as in (c).
In (d-f) the dashed vertical line indicates the corresponding value of $\kappa$.}
\label{fig5}
\end{center}
\end{figure*}

In Figs. \ref{fig5}(a-c) the cluster distribution $N_c(n)$ for three different cases are plotted as a function
of cluster size $n$. The distributions are approximately exponential with a decay parameter
$\lambda$ that depends on aspect ratio and packing fraction. 
The exponential behaviour of $N_c(n)$ was already observed in Monte Carlo simulations of thermally fluctuating 
hard rectangles \cite{PRE}, and can be attributed to a mechanism for cluster formation where the probability of a 
particle joining (or leaving) a cluster is approximately independent of the cluster size \cite{frag1}.

\begin{figure}
\begin{center}
\includegraphics[width=0.95\linewidth,angle=0]{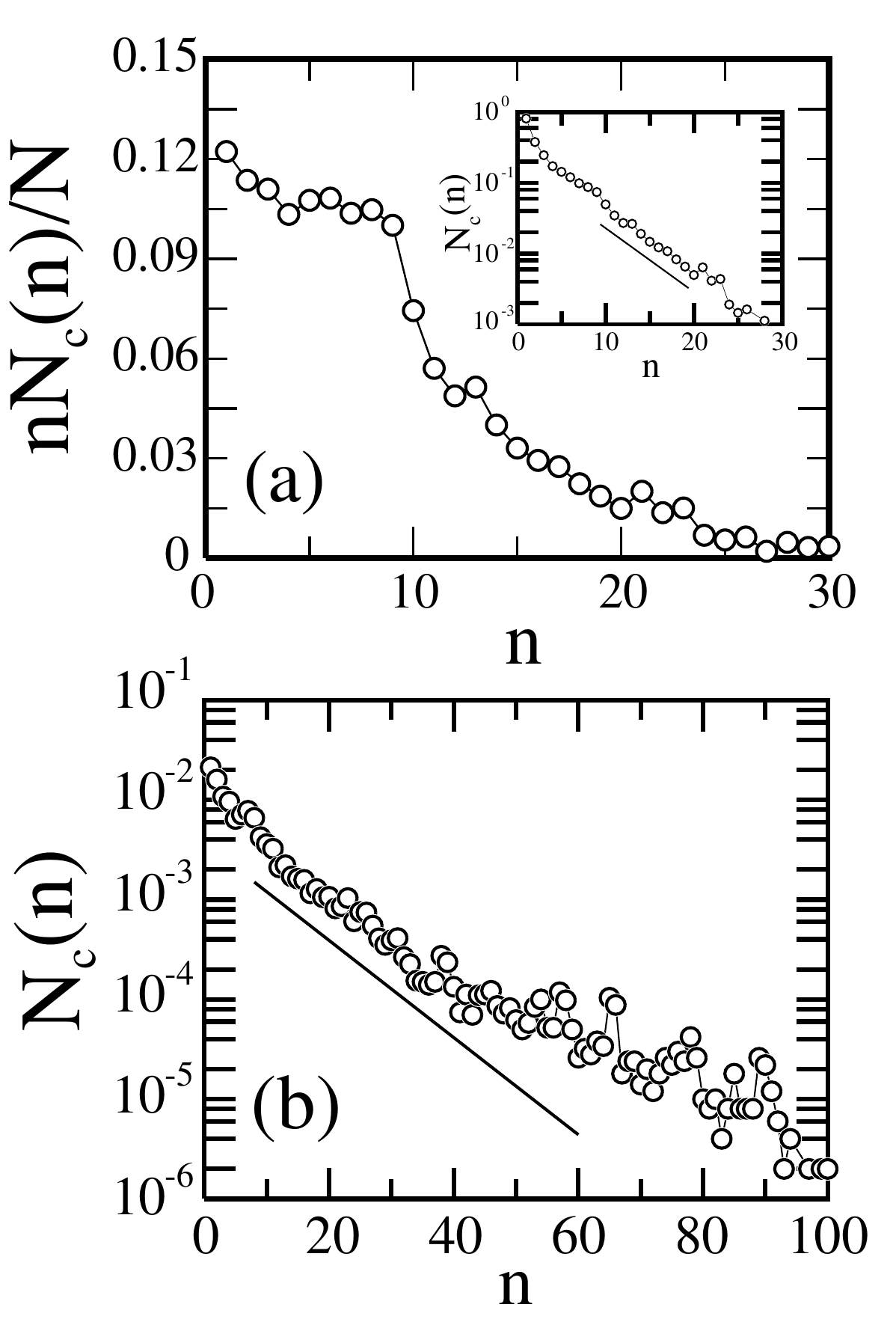}
\caption{(a) Fraction of particles in clusters of a given size, $nN_c(n)/N$, as a function of cluster size $n$, 
for the case $\kappa=10$ and $\varphi=0.64$. The inset is a semilogarithmic graph.
(b) Normalised cluster distribution functions $N_c(n)$ as a function of cluster size $n$ for the case
$\kappa=8$ and $\varphi=0.81$. The straight line in both panels corresponds to exponential behaviour.}
\label{dist_large}
\end{center}
\end{figure}

Figs. \ref{fig5}(d-f) show the distribution $nN_c(n)$ for the same cases as in panels (a-c). These distributions 
give the fraction of particles in the system that belong to a 
cluster of size $n$. For reference, we have drawn in each case a vertical dashed line located at a cluster size equal
to the corresponding aspect ratio.
Note that the position of these lines are very close to the maximum of the corresponding distributions. This feature 
is general for $\kappa\alt 8$, and means that particles have some preference to arrange into square clusters
with size $n\simeq\kappa$. The effect is not so clear for $\kappa > 8$, see Fig. \ref{dist_large}(a), although a change
in behaviour clearly exists at $n\simeq\kappa$. 
Our conclusion is that, for particles of low aspect ratio, the probability of a particle 
to be part of a cluster of size $n$ is maximum when $n\simeq\kappa$, i.e. for close-to-square 
clusters. These clusters have a characteristic axis (along or perpendicular to the particle long axes, according to whether
$n<\kappa$ or $n>\kappa$, respectively) and seem to be particularly stable during the course 
of the experiment.
Interaction between these clusters drives the formation of extended ordered patterns in the system, as will be seen below.

On closer inspection, some cluster distributions are seen to be more complex than a simple exponential. First, for
sizes $n\alt\kappa$, the distribution is usually not exponential. Second, and more significant, is the fact that, as packing fraction 
is increased, the distributions develop a nonexponential tail at larger and larger values of $n$. At high $\varphi$, they
correspond to fat-tailed distributions, rather than to pure-exponential,
indicating the existence of a marked process of cluster aggregation. An example is given in 
Fig. \ref{dist_large}(b). This feature signals the formation of large ordered structures that will be analysed below.

\begin{figure}
\begin{center}
\includegraphics[width=1.00\linewidth,angle=0]{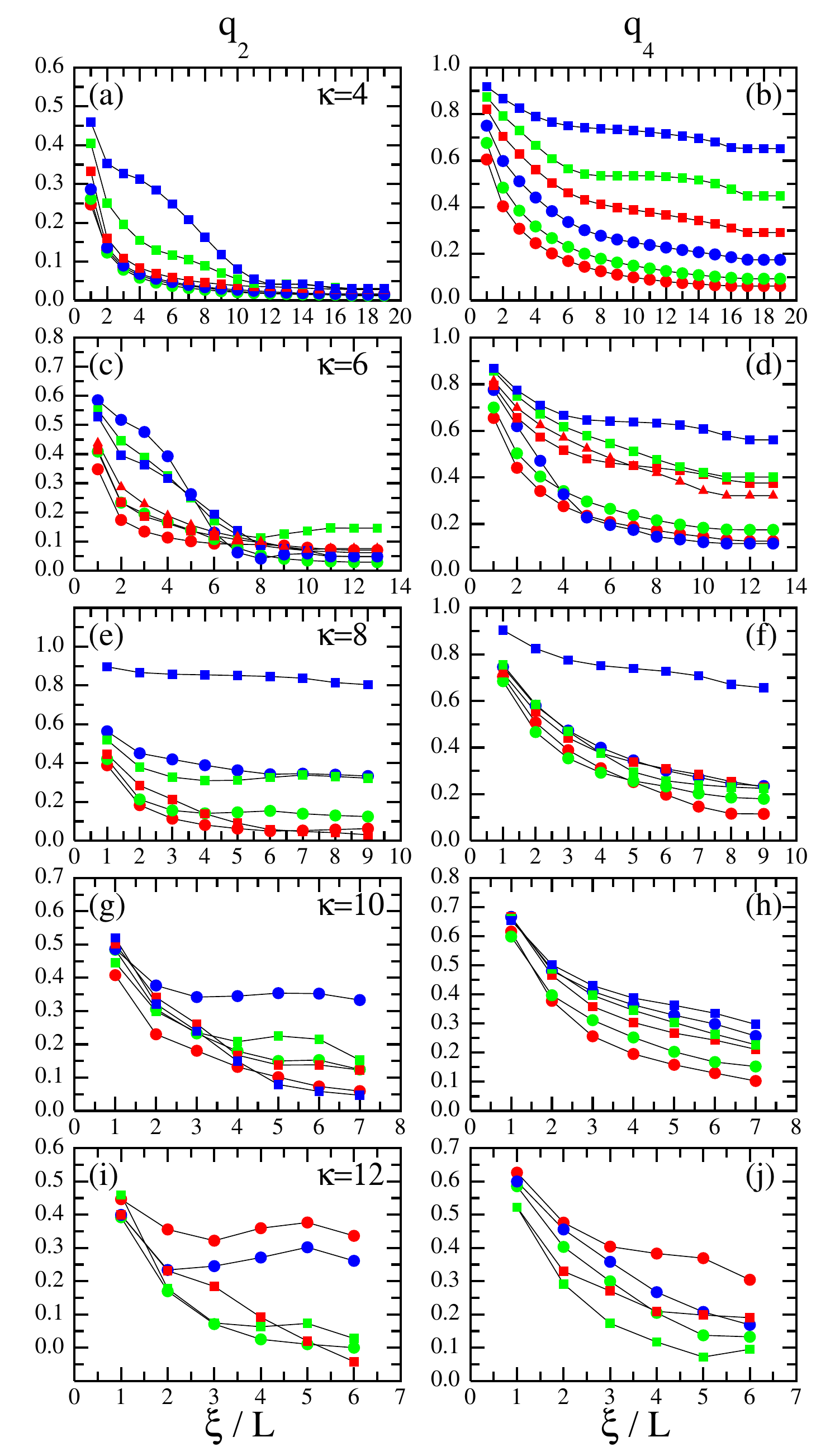}
\caption{Dependence of order parameters $q_2$ and $q_4$ on radius of averaging circular region $\xi/L$, scaled with particle length $L$. 
Left panels (a), (c), (e), (g) and (i) refer to $q_2$, while right panels (b), (d), (f), (h) and (j) depict $q_4$. Aspect ratio $\kappa$ 
increases from top to bottom, as indicated by labels. Different symbols indicate different densities, in the sequence 
\tikzcircle[white, fill=red]{3.5pt}, \tikzcircle[white, fill=green]{3.5pt}, \tikzcircle[white, fill=blue]{3.5pt}, 
\tikzsquare[white, fill=red]{6.5pt},\hspace{0.1cm}\tikzsquare[white, fill=green]{6.5pt},
\hspace{0.0cm}\tikzsquare[white, fill=blue]{6.5pt} and 
\textcolor{myred}{$\blacktriangle$}
for increasing density. For each value of $\kappa$, the sequences in $\varphi$ are: for $\kappa=4$, $0.52$, 
$0.59$, $0.64$, $0.68$, $0.70$ and $0.75$; for $\kappa=6$, $0.55$, $0.60$, $0.64$, $0.69$, $0.74$, $0.78$ and $0.80$;
for $\kappa=8$, $0.65$, $0.67$, $0.70$, $0.74$, $0.78$ and $0.81$; for $\kappa=10$, $0.59$, $0.64$, $0.70$, $0.73$, $0.78$ and 
$0.82$; and for $\kappa=12$, $0.65$, $0.69$, $0.73$, $0.77$ and $0.81$.  
}
\label{op_local}
\end{center}
\end{figure}

\begin{figure}
\begin{center}
\includegraphics[width=0.95\linewidth,angle=0]{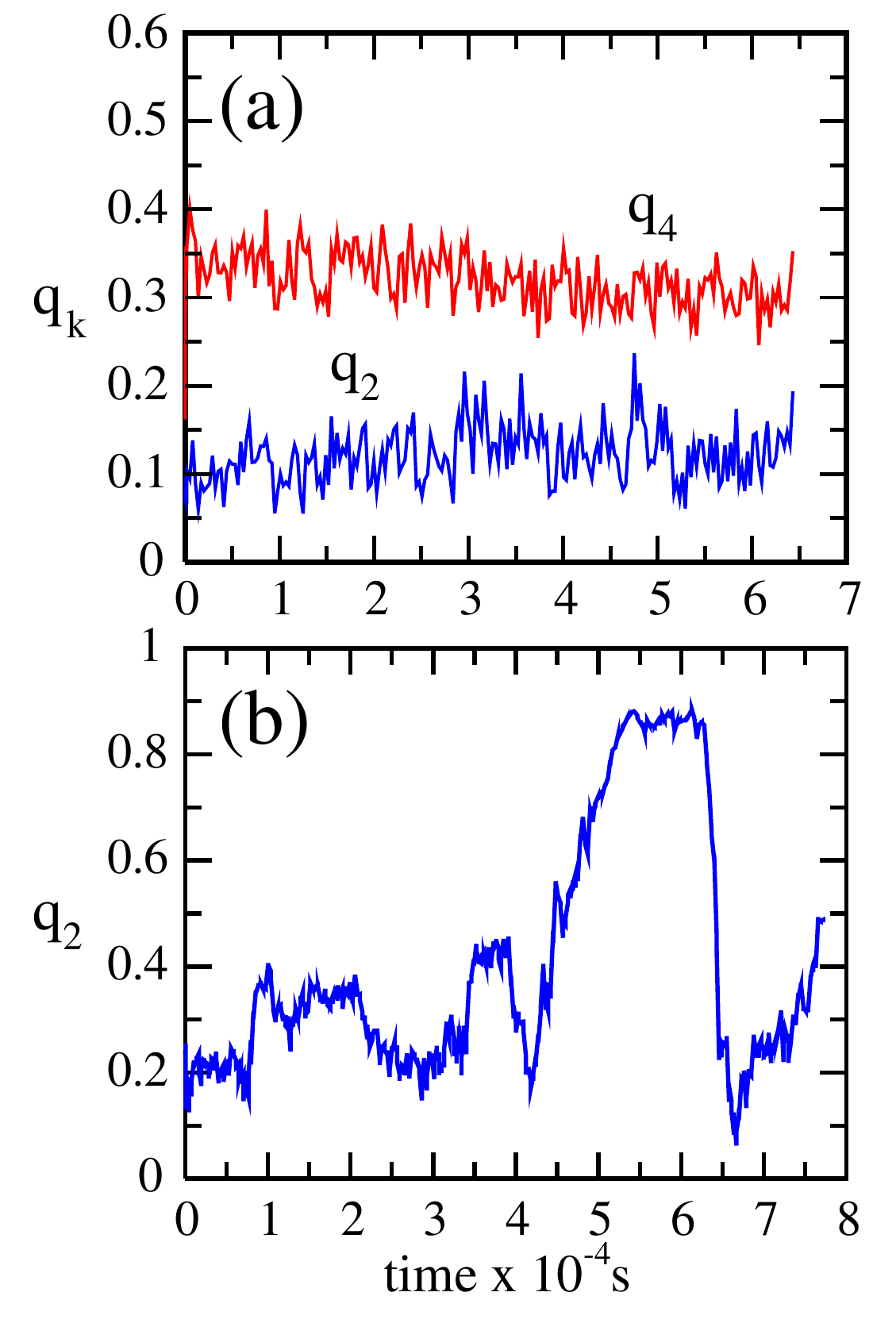}
\caption{(a) Time evolution of the uniaxial $q_2$ and tetratic $q_4$ order parameters for an experiment
with $\kappa=4$ and $\varphi=0.64$. (b) Time evolution of uniaxial nematic order parameter $q_2$ for the case
$\kappa=8$ and $\varphi=0.81$. In this experiment the system exhibits local smectic fluctuations.}
\label{fig_time}
\end{center}
\end{figure}

\subsection{Order parameters and dynamical evolution}

In this section we look at order parameters and angular distribution functions. These results are crucial to
identify the type of ordering in the system \cite{daniel}. Due to spatial and time fluctuations, the order parameters
depend on both position and time, and proper local fields $q_2({\bm r},t)$ and $q_4({\bm r},t)$ have to be obtained from
the particle positions. To obtain the fields at some point ${\bm r}$, we average over the particles located within a circular
region of radius $\xi$ centred at ${\bm r}$. The values of $q_k({\bm r},t;\xi)$ depend on the radius $\xi$; one would expect 
these values to saturate for a size $\xi$ of the order of the coherence length or typical domain size. This is approximately
the case, but it is not possible to establish a clearcut criterion. Fig. \ref{op_local} shows the time-averaged values of 
the local parameters, as a function of $\xi/L$, for all cases studied. In general the order parameters decrease with $\xi$. From 
the figures, we have reached a compromise and in the following we set the values for the radius used to calculate the local 
order parameters to $\xi=4L$.

\begin{figure}
\begin{center}
\includegraphics[width=0.95\linewidth,angle=0]{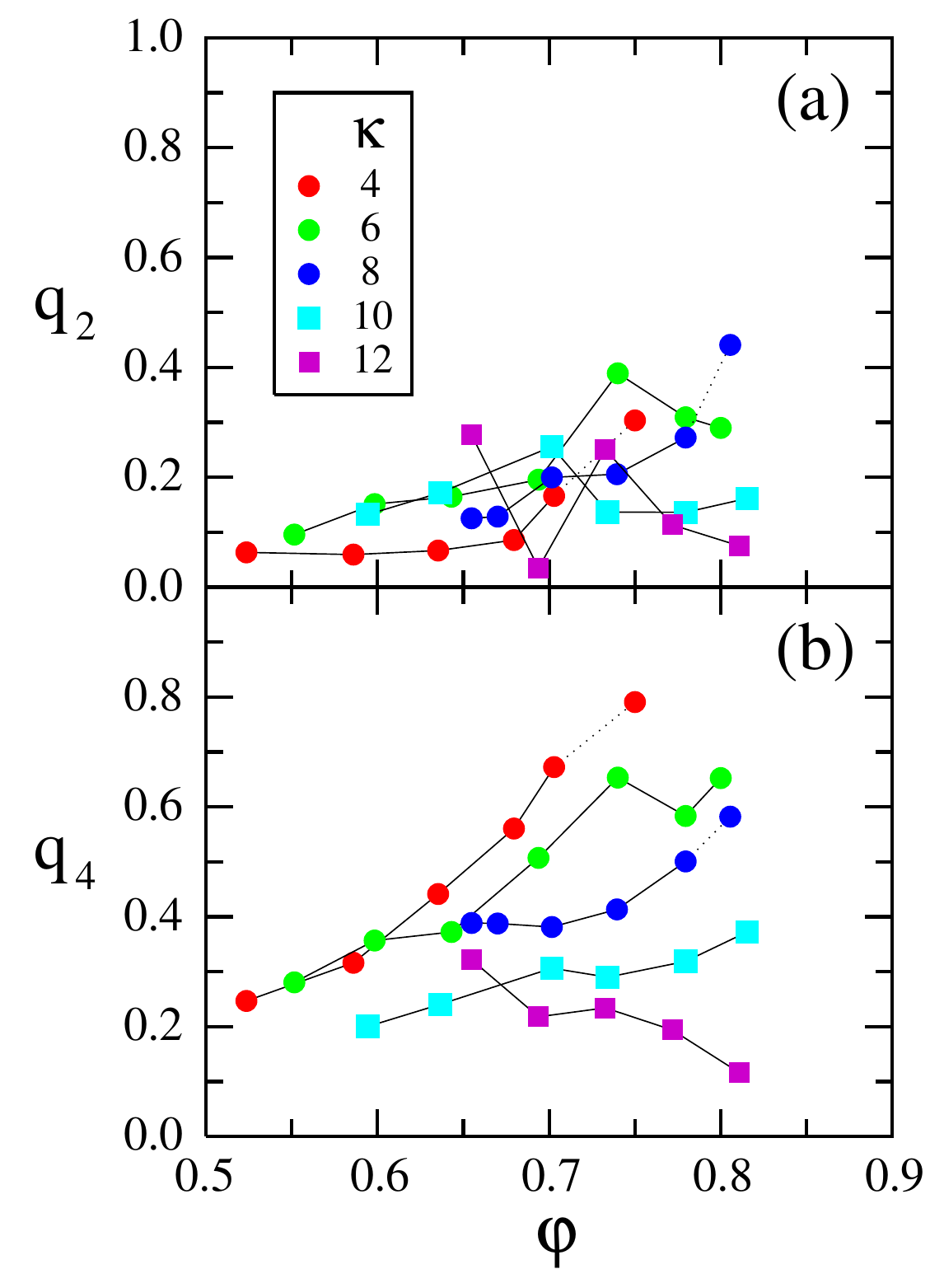}
\caption{(a) Uniaxial nematic $q_2$ and (b) tetratic $q_4$ order parameters as a function of particle packing fraction $\varphi$ for different aspect ratios $\kappa$
(indicated in the keybox). The dotted lines indicate that the system exhibits large smectic fluctuations. In this case values were obtained from averages over the whole experiment.}
\label{h1}
\end{center}
\end{figure}

Let us now examine the time evolution of the order parameters.
Fig. \ref{fig_time}(a) shows the time evolution of the order parameters $q_2$ and $q_4$ for an experiment with $\kappa=6$ and $\varphi=0.65$. 
After a short transient of approximately $10^{3}$ s, the order
parameters level off at more or less constant values with large fluctuations about these values, of typically
$20\%$ with respect to the mean. These steady-state values of the order parameters correspond to a well-developed
and stable tetratic configuration. Note that a purely tetratic state would have $q_2=0$, $q_4\ne 0$. In our case,
where the number of particles is $N\sim 10^3$, we expect a non-zero value of $q_2$ due to the finite number of
particles. 

\begin{figure*}
\begin{center}
\includegraphics[width=1.0\linewidth,angle=0]{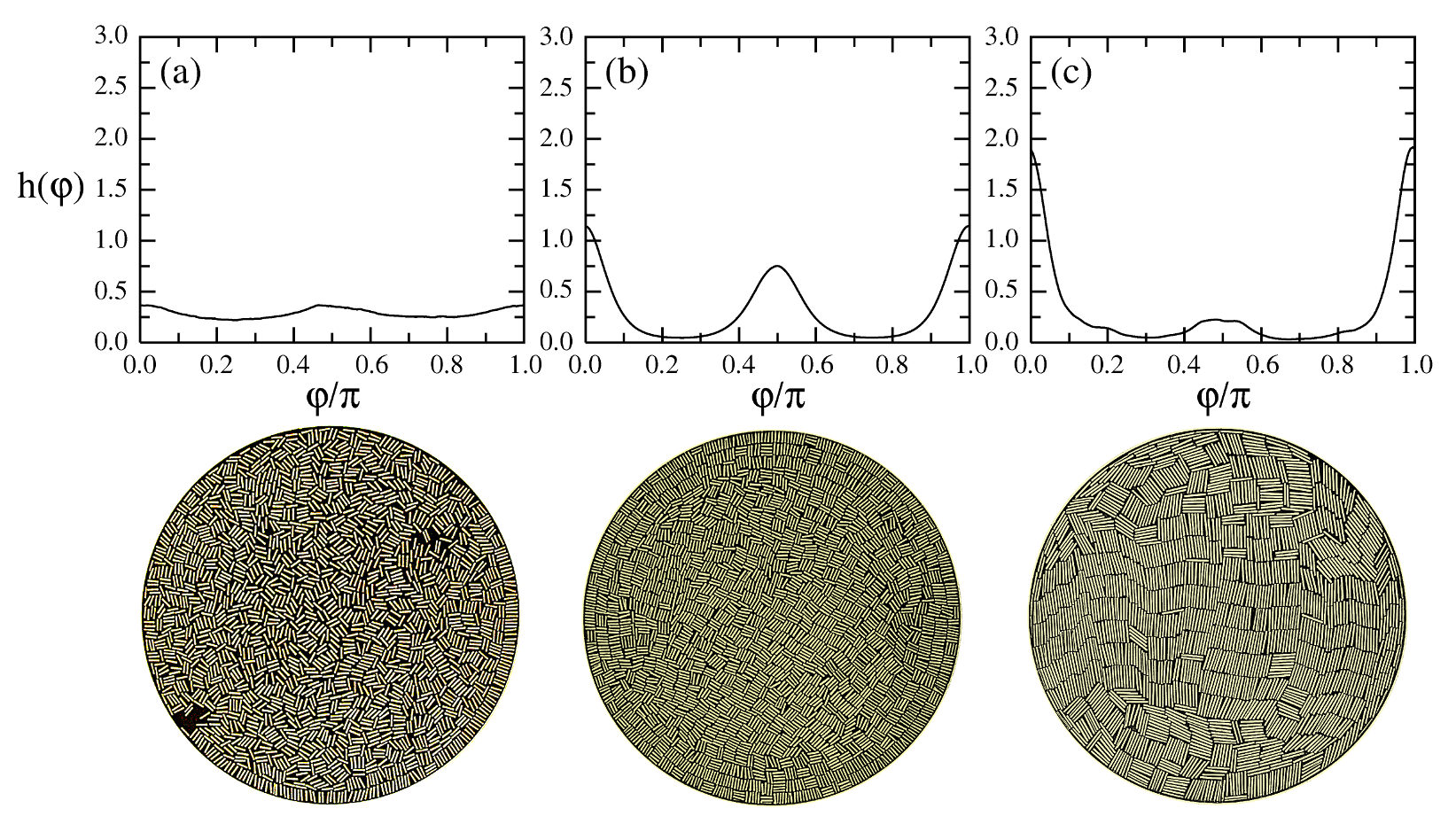}
\caption{Orientational distribution functions for three different cases.
(a) $\kappa=4$ and $\varphi=0.52$, isotropic configuration. (b) $\kappa=4$ and $\varphi=0.70$,
configuration with large tetratic domains. (c) $\kappa=8$ and $\varphi=0.81$, configuration with large smectic domains.
Typical snapshots are shown in each case.}
\label{distributions}
\end{center}
\end{figure*}

However, we have identified cases where order parameters fluctuate strongly, and a
steady state does not seem to exist. These cases always correspond to high-density configurations, irrespective
of the aspect ratio. An example is given in Fig. \ref{fig_time}(b), where the time evolution of $q_2$
for the case $\kappa=8$ and $\varphi=0.81$ is shown. Orientational order builds up with time, but there
seem to be abrupt events in the system which destroy order. We later argue that these situations are associated
to the formation of smectic fluctuations at high density. Smectic domains are very sensitive to hydrodynamic
shear modes, which are known to be excited in vibrated monolayers of rods \cite{Aranson}.
Note that the duration of the experiments in Figs. \ref{fig_time}(a) and (b)
was particularly long since the aim
was to establish the existence of steady states in the system. With this information,
the duration of the experiments to be reported in the following were adjusted 
according to the time evolution of the order parameters. 

For intermediate values of $\varphi$, steady-state values for $q_2$ and $q_4$ can always be obtained. 
Fig. \ref{h1} collects the results for
these order parameters from experiments with particles of different aspect ratios $\kappa$. We can identify two 
general trends:
(i) the uniaxial order parameter $q_2$ is low
(ii) the tetratic order parameter 
increases steadily with packing fraction for particles with $\kappa=4$ and $6$, while it remains constant for $10$ and $12$, 
the case $8$ being a critical case. 
At even higher packing fractions the $q_2$ order parameter rises up to relatively
high values; this is due to the appearance of smectic domains in the system. Smectic fluctuations will be covered in
more detail later. Overall we can say that there are no sharp boundaries between the differently ordered states, and that
large fluctuations or domains with one type of order may be found in samples with a predominance of a different type of 
order in large intervals of packing fraction. This observation is at variance with the findings in Ref. \cite{daniel},
where a detailed `phase diagram' including sharp `phase transitions' was obtained. However, we must be cautious when
emphasising the different results of the experiments, since
different protocols to quantify the order may lead to different conclusions on the global ordering behaviour.

To more easily connect the order parameters with particle configurations, Fig. \ref{distributions} 
shows distribution functions for the cases $\kappa=4$ and $8$ and different packing fractions
in each case, along with typical snapshots.
Panel (a) corresponds to the case $\kappa=4$ and $\varphi=0.52$. The orientation distribution is
uniform and the steady-state configurations pertain to the isotropic phase. Panel (b) corresponds to the same
aspect ratio, but the packing fraction is increased to $\varphi=0.70$. The distribution exhibits two peaks at
$\phi=0^{\circ}$ and $90^{\circ}$, which corresponds to configurations containing large domains with tetratic ordering.
Note in the image the considerable layering next to the circular wall, extending up to three layers into the cavity.
Finally, panel (c) corresponds to the case $\kappa=8$ and $\varphi=0.81$. Here configurations containing
large uniaxial smectic domains, with a much lower secondary maximum in the
orientation distributions, are formed; this case corresponds to the high value of $q_2$ visible in Fig. \ref{h1}. 
We note that, in (a), the distribution function is not completely flat, but
presents some inherent structure. This is due to a bias introduced by our procedure to translate the origin of angles
to the direction given by the director in each sampling region. In essence this effect gives rise to non-zero values
of the order parameters for regions containing a few non-oriented rectangles.

One important result of our experiments is that it was not possible to identify truly uniaxial nematic configurations,
not even locally, for all particle aspect ratios explored. This result includes the longer particles, 
for which both equilibrium systems and the vibration experiments of Ref. \cite{daniel} predict uniaxial nematic
phases. This phenomenon is signalled by the fact that the local
uniaxial nematic order parameter, $q_2$, is always low, regardless of the size chosen for the region where
the local average is performed. This seems to indicate that uniaxial correlations in the direction of the long axis of a particle
decay very fast (except in the case where well-developed smectic fluctuations are present), precluding the
formation of local uniaxial nematic order. We think that
this behaviour is a consequence of the strong clustering tendency
exhibited by a granular system of rectangular particles.
Uniaxial nematic phases have been identified by Narayan and coworkers in vibrated monolayers of steel rolling pins
\cite{hindues}, but cylinders only exhibited tetratic order. However, in Ref. \cite{daniel} uniaxial nematic phases
were identified in systems of plastic cylinders. We have used the same plastic cylinders in our vibration setup and
confirm our results for steel particles. However, it is true that the experimental conditions of
the two experiments (free particle height, cavity size, etc.) are not identical, and therefore
the different results are not conclusive.
The identification of ordered phases may be a
subtle question in these dissipative granular systems subject to fluctuations and stringent geometrical constraints,
and further studies on this question are desirable. 

\begin{figure}
\begin{center}
\includegraphics[width=0.95\linewidth,angle=0]{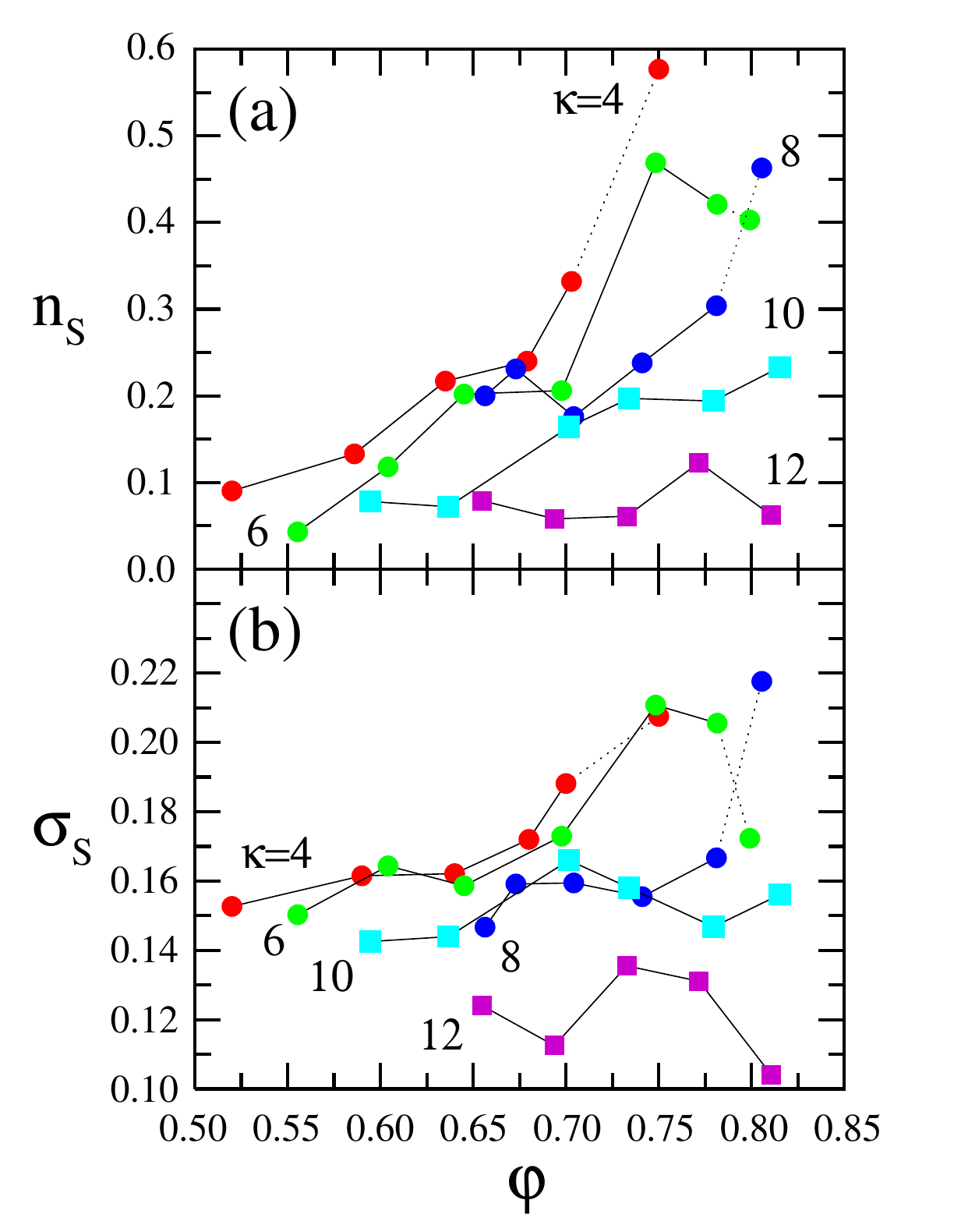}
\caption{Two measures of smectic order as a function of packing fraction.
(a) $n_S$, which is the fraction of particles belonging to
clusters of size larger than $n_c=2\kappa$. (b) Smectic order parameter $\sigma_s$.
In both cases data are provided for all the aspect ratios $\kappa$ explored
(indicated by labels). The dotted lines indicate that the system exhibits large smectic fluctuations.}
\label{smec3}
\end{center}
\end{figure}

In the range of packing fractions and particle aspect ratios where M\"uller et al. \cite{daniel} claim to find a uniaxial
nematic phase, we have observed instead a random mixture of clusters or `patchy' state consisting of large groups of 
particles in parallel arrangements, with groups oriented more or less at random with respect to each other. An example
of this state is given in Fig. \ref{fig1}. As already 
discussed, these groups of particles are quite stable, since they have a long lifetime compared with the
typical diffusion time (obviously the identity of a given cluster remains
the same provided only a few particles are joining or leaving the cluster in a short time). The example of Fig. \ref{fig1} corresponds to the uniaxial nematic region of Ref. \cite{daniel} (note
that the effective cavity sizes of the two experiments are equivalent). Clearly local tetratic order is absent, but
so is local uniaxial nematic ordering. Again, different experimental conditions may be causing different behaviours.

\begin{figure*}
\begin{center}
\includegraphics[width=1.00\linewidth,angle=0]{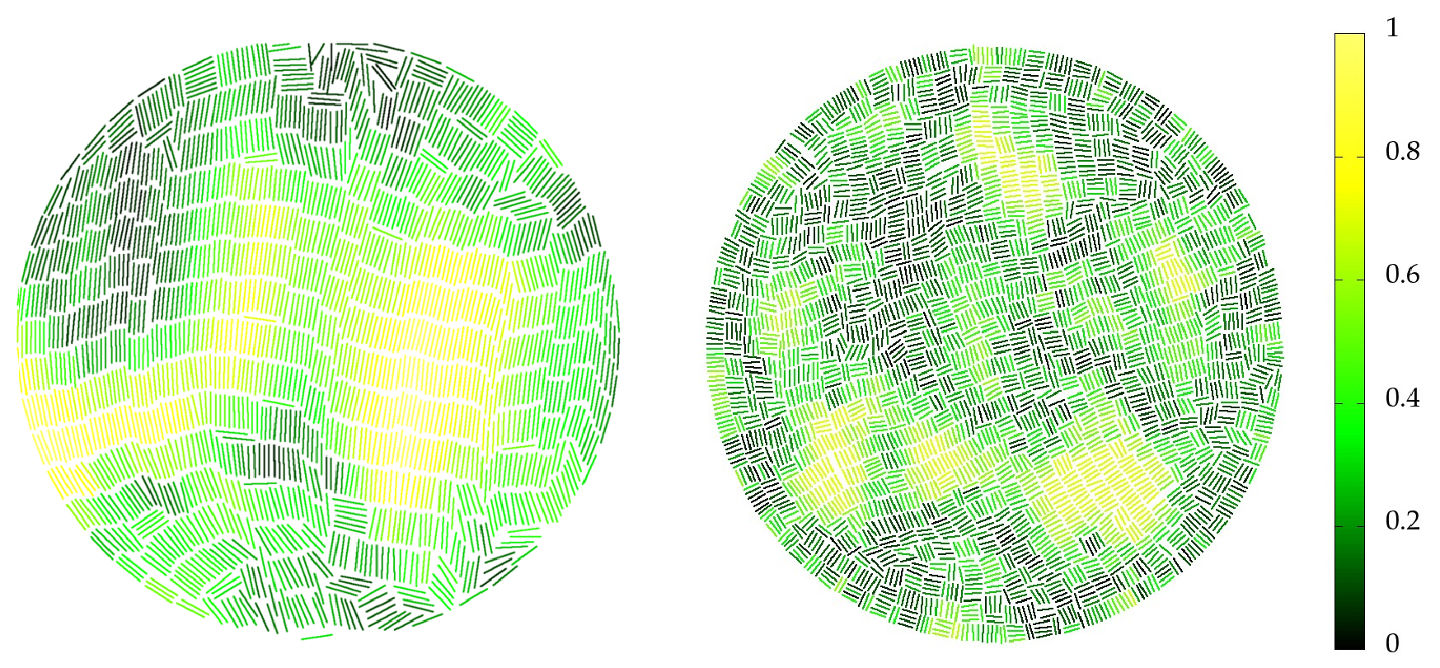}
\caption{Colour maps of the local smectic order parameter $\sigma_S$ for two cases.
Left panel: $\kappa=8$ and $\varphi=0.81$. Right panel: $\kappa=4$ and $\varphi=0.75$. The vertical bar indicates
the scale of the parameter.
Note that that particles have been shortened by a small amount
to improve visualization. Holes correspond to particles that could not be identified by the imaging software.}
\label{smec}
\end{center}
\end{figure*}

\begin{figure}
\begin{center}
\includegraphics[width=1.05\linewidth,angle=0]{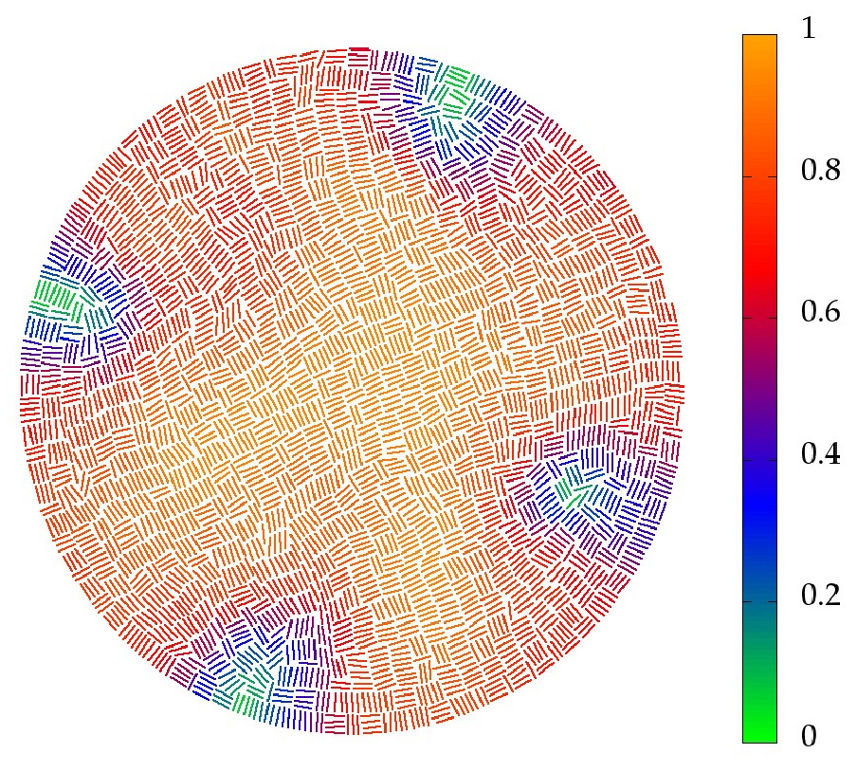}
\caption{Colour map of the local tetratic order parameter $q_4$ for the case
$\kappa=4$ and $\varphi=0.75$.
Note that particles have been shortened by a small amount
to improve visualization. Holes correspond to particles that could not be identified by the imaging software.}
\label{tetratic_map}
\end{center}
\end{figure}

Summing up our findings on orientational order in the monolayer, Fig. \ref{op_local} reflects the behaviour in a very transparent way.
For $\kappa=4$ and $6$ $q_2$ is low and $q_4$ high, a condition for tetratic ordering. At high density $q_2$ begins to increase at sizes $r/L$ compatible
with the observed fluctuating smectic domains. $\kappa=10$ and $12$ exhibit low values of both order parameters, corresponding to the patchy state
with locally frustated orientational order; at higher densities the order parameters increase, again reflecting the formation of small-size 
smectic fluctuations. $\kappa=8$ seems to be a critical case.

Let us now turn to the spatially-ordered, smectic fluctuations.
Fig. \ref{smec3} shows the changes in the smectic parameters, defined in the Appendix,
as a function of packing fraction, for the different aspect ratios explored. $\sigma_S$ is the
standard smectic order parameter, while $n_S$ reflects the amount of particles belonging to large clusters.
Overall both parameters become larger as density increases. Their evolution
is highly correlated, but $n_S$ gives a clearer picture. Again, it is not
possible to identify a sharp transition or at least a region where smectic
order beings to increase at a significantly high ratio. In an effort to understand
the gradual and rather smooth increase in smecticity with density, we looked
at videos of the time evolution of the system (see Supplemental 
Material \cite{videos}). Obviously uniform
smectic order is hampered by the circular geometry of the cavity, which is incompatible
with the uniaxial symmetry of the smectic phase and should induce the presence of defects;
this is not observed in our system. From a close
examination of the videos, one can conclude
that the time evolution of these systems at high packing fraction is very dynamic,
with smectic domains forming, living for relatively short times, disappearing and reappearing at
later times. An example of a smectic fluctuation is given in Fig. \ref{smec}, where a false-colour map
of a field of the parameter $\sigma_S$ (obtained by averaging over a short time interval) is superimposed on a configuration
at a corresponding instant of time. The size of smectic 
domains increases with density. 

To support this conclusion, Fig. \ref{fig_time}(b) shows a time evolution
of the parameter $q_2$ for an experiment with $\kappa=8$ and $\varphi=0.81$.
In this experiment the system exhibits large smectic-domain fluctuations,
which are clearly correlated with the uniaxial order parameter $q_2$.
Time variations of the smectic order parameter $\sigma_S$ and the parameter
$n_S$ are perfectly correlated with those of $q_2$ (not shown).
From the figure, it is
apparent that smectic order can reach high values, but there are
repeated, abrupt falls in $q_2$ with time. We believe that these sudden falls, after which
the order parameter rises up again, are related with the increase of
hydrodynamic flow and vorticity \cite{Aranson}.
Smecticity and vorticity are present, one suppressing the other, during a given experiment. 
By contrast, tetratic order can develop uniformly 
in the cavity, as demonstrated in Fig. \ref{tetratic_map}, where the
false-colour map represents the local field $q_4$, again 
supersimposed on a configuration at a corresponding instant of time.
To restore the symmetry of the phase and at the same time satisfy the 
preferred orientation at the walls the system develops four point defects
symmetrically located at relative angles of $90^{\circ}$.
This proves that, when the velocity flow is suppressed, the granular monolayer 
behaves as a liquid-crystalline material in thermal equilibrium, whose
properties can be understood from a competition between surface alignment,
elasticity and defects that restore the symmetry \cite{galanis2}.

\begin{figure}
\begin{center}
\includegraphics[width=1.10\linewidth,angle=0]{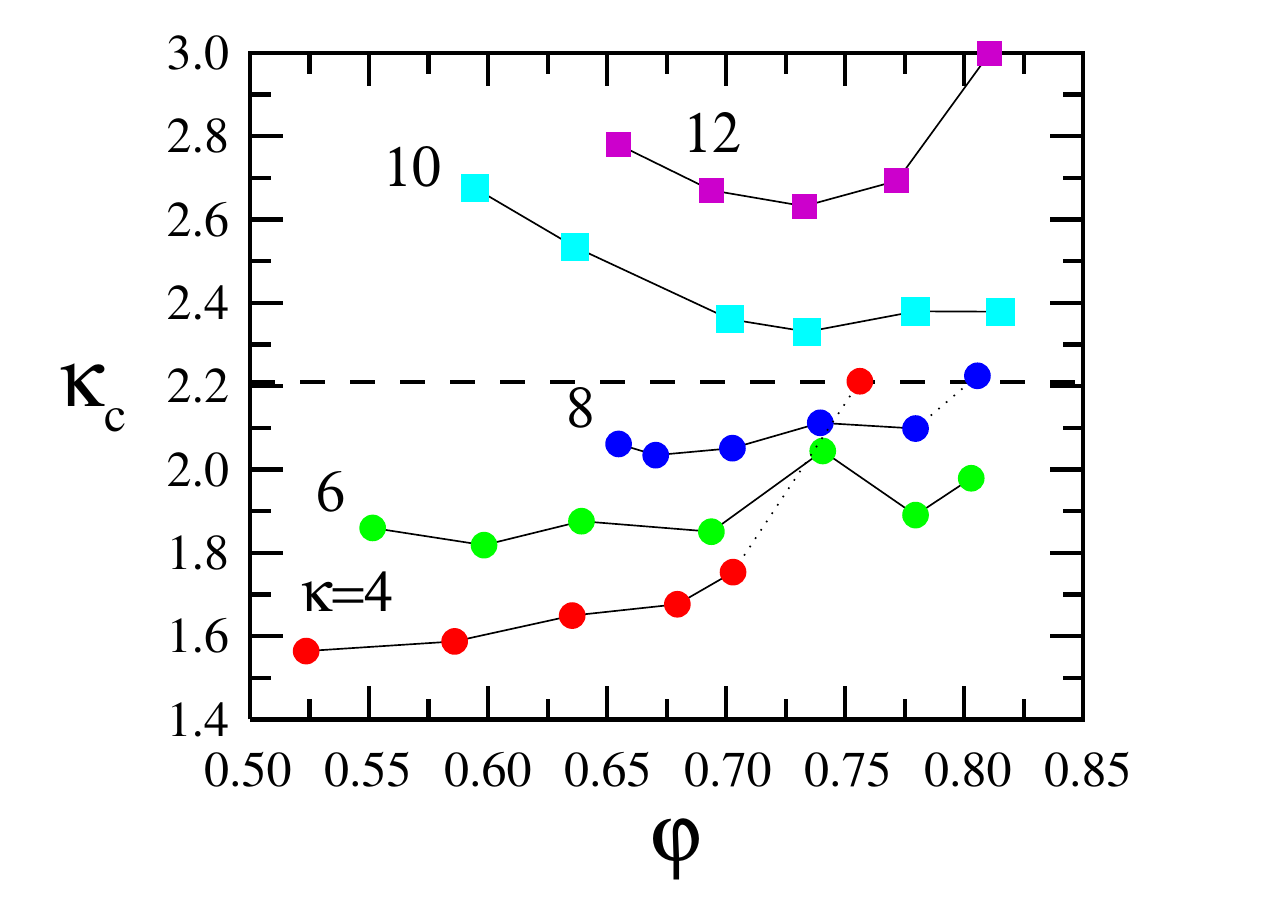}
\caption{\label{kappa_media} Mean cluster aspect ratio as a function of packing fraction for the aspect ratios explored
(indicated as labels). The dotted lines indicate that the system exhibits large smectic fluctuations. The dashed horizontal
line indicates the critical value of aspect ratio for tetratic stability according to scaled-particle theory \cite{SPT}.}
\end{center}
\end{figure} 

\begin{figure}
\begin{center}
\includegraphics[width=1.05\linewidth,angle=0]{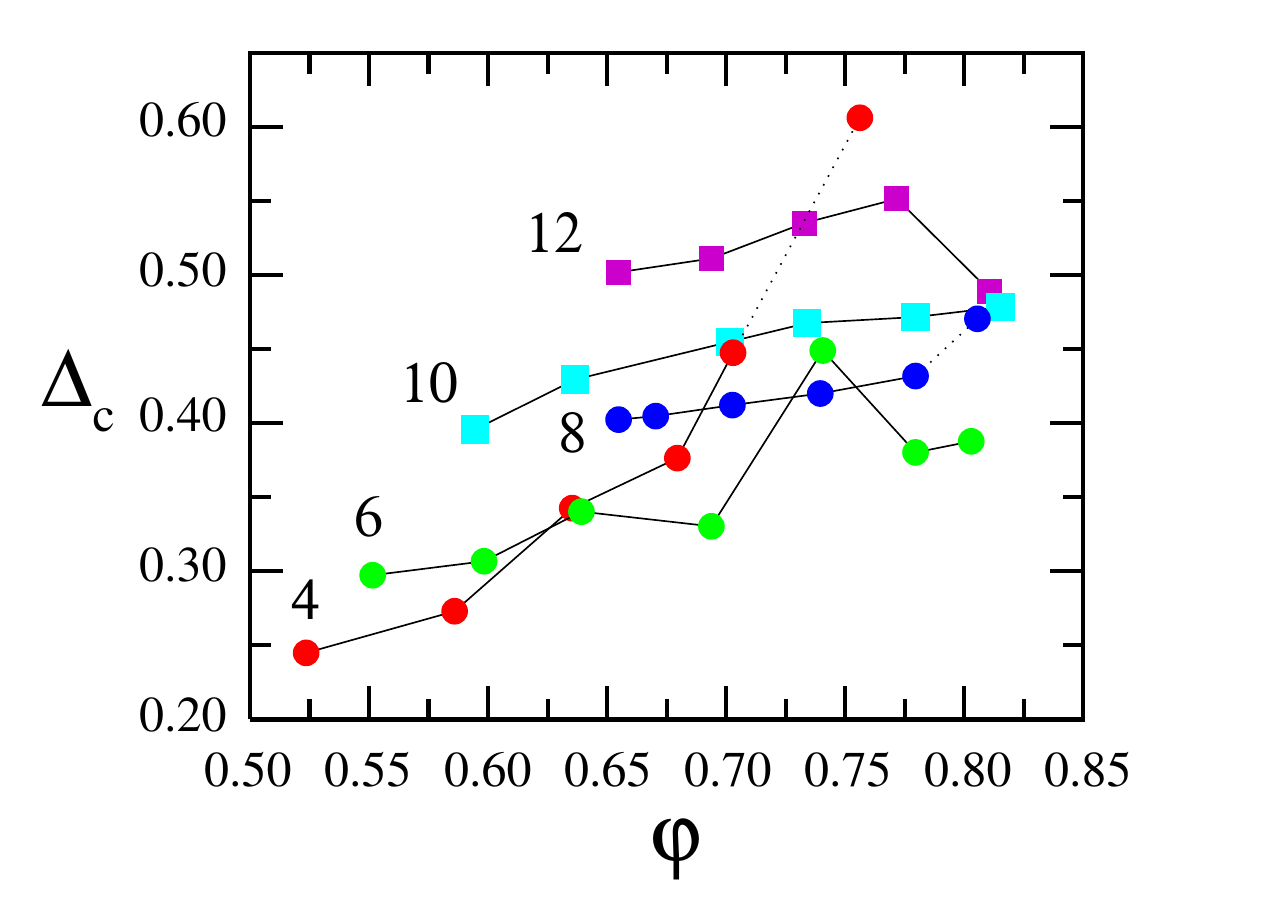}
\caption{\label{kappa_poly} Polydispersity of the cluster aspect ratio as a function of packing fraction 
for the aspect ratios explored (indicated as labels). The dotted lines indicate that the system exhibits large smectic fluctuations.}
\end{center}
\end{figure}

\begin{figure}
\begin{center}
\includegraphics[width=0.95\linewidth,angle=0]{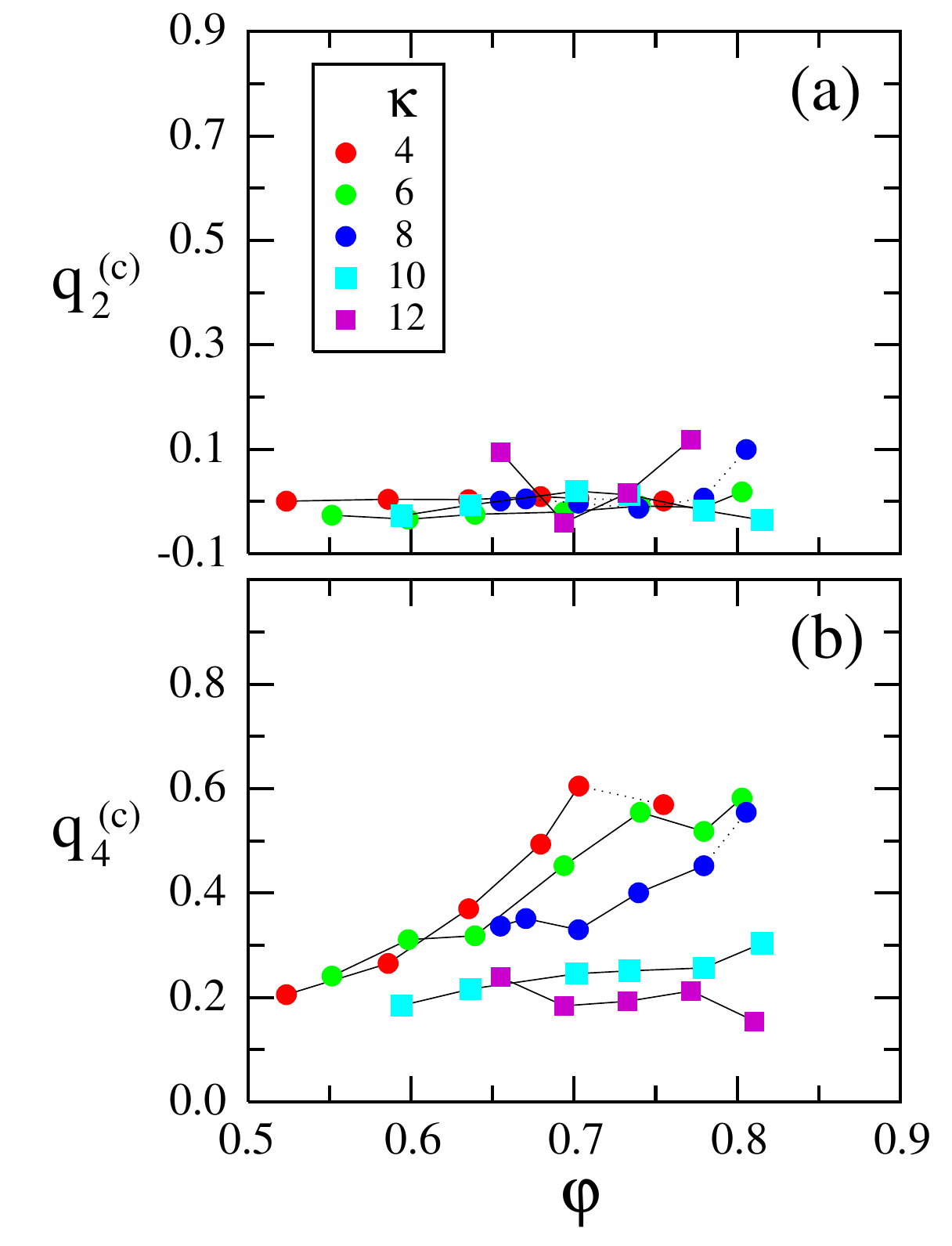}
\caption{\label{qc} Order parameters $q_2$ and $q_4$ for the mixture of superparticles as a function of
packing fraction and for the different aspect ratios explored (indicated in the keybox).
The dotted lines indicate that the system exhibits large smectic fluctuations.}
\end{center}
\end{figure}

\subsection{Cluster-mixture view}

Based on the high stability of clusters, it is tempting to regard the system as a collection of 
superparticles with their own ordering properties. To look at this picture in more detail, we have 
computed angular distribution functions and corresponding order parameters for the superparticles. 
Some interesting results emerge from this analysis. For example, one can calculate an effective mean and dispersion in the 
aspect ratio of clusters,
\begin{eqnarray}
\kappa_c=\left<\frac{L_c}{D_c}\right>_i,\hspace{1cm}\Delta_c=\frac{1}{\kappa_c}\sqrt{
\left<\left(\frac{L_c}{D_c}\right)^2-\kappa_c^2\right>_i},
\end{eqnarray}
where the length and width of a cluster is $L_c=\hbox{max}(L,\delta)$, $D_c=\hbox{min}(L,\delta)$, with $\delta$ the maximum centre-to-centre
distance between two particles in the cluster, and where index $i$ runs over all the clusters detected in the stationary regime of the
experiment.

Fig. \ref{kappa_media} shows the behaviour of 
$\kappa_c$ as a function of $\varphi$ for all the particles aspect ratios $\kappa$ analysed. One observes that,
for a particular $\kappa$, the effective mean cluster ratio is remarkably insensitive to $\varphi$.
The dispersion $\Delta_c$ (which is close to the cluster length 
polydispersity) is more sensitive to $\varphi$, see Fig. \ref{kappa_poly}. All of these results suggest that, for a given
$\kappa$, the system can be regarded as a mixture of polydisperse superparticles of rather constant aspect ratio and a 
relatively high polydispersity that increases with density. But still more interesting is the fact that, in all cases,
$\kappa_c<\kappa$, and that: (i) systems forming tetratic states ($\kappa=4$ and $6$) have $\kappa_c\alt 2$;
(ii) systems not forming tetratic states ($\kappa=10$ and $12$) have $\kappa_c\agt 2.3$; and (iii) the critical
system $\kappa=8$ has $2.0\alt\kappa_c\alt 2.2$. Therefore, $\kappa^*\simeq 2.1$ can be regarded as a critical aspect ratio
for the formation of tetratic ordering. This is surprisingly close to the critical value predicted by scaled-particle theory,
which gives $\kappa^*=2.2$ \cite{SPT}. It is tempting to suggest that the strong clustering effects in the vibrated system somehow
renormalise the interaction units from rectangles to clusters or superparticles. Since most of the important short-range 
particle correlations are `hidden' inside the superparticles, the remaining weak correlations between superparticles 
can be renormalised into effective two-particle (excluded-area) correlations, well accounted for by standard mean-field theories.
A further result that may support this idea is the behaviour of the cluster order parameters, $q_2^{(c)}$ and $q_4^{(c)}$, 
presented in Fig. \ref{qc}. This behaviour is remarkably similar to the one shown by the order parameters of particles,
Fig. \ref{h1}.

\section{Discussion and conclusions}
\label{discussion}

The results presented in the previous section point to a scenario for the ordering properties of vibrated rods 
a bit more complicated than previously reported \cite{daniel}. On the one hand, we could not identify a sharp transition point between
the disordered isotropic phase and the orientationally order phases at any value of aspect ratio. Rather, orientational
order is built up in a continuous fashion as packing fraction is increased, and probably finite size and boundary effects also 
contribute to the smearing down of the transitions. On the other hand, the oriented configuration of the system is different for aspect
ratios $\kappa<8$ and $\kappa>8$. This is consistent with the critical value $\kappa^*=7.3$ found in Ref. \cite{daniel},
which separates tetratic from uniaxial-nematic ordering. But the nature of our phases is not completely identical.
In the first case, $\kappa<8$, clear tetratic ordering is observed, as in \cite{daniel}. But in the second, $\kappa>8$, the system forms
large clusters of parallel particles with a random relative orientation, and uniaxial nematic configurations cannot be defined, even locally.
At large packing fractions all systems form smectic domains, whose size and life-time increase with packing fraction. Smectic order,
which can be understood as arising from packing maximisation,
competes with the formation of vortices in the system, and uniform, 
well-defined steady states apparently do not exist. 

The formation of ordered structures in the system may also be viewed in terms of particle clustering. The tendency
of dissipative rods to cluster is much stronger than in thermal equilibrium, and the cluster concept becomes more useful. For low aspect
ratio, particles organise into clusters with a close-to-square shape, low shape polydispersity and high rotational mobility. These
factors favour local configurations with tetratic order where neighbouring clusters are oriented at orthogonal directions. 
Clusters with low aspect ratio form easily, because they involve a small number 
of rods and consequently recruit low-order particle correlations. By contrast, for long rods, 
the system forms compact clusters of large and uneven sizes that are curved and cannot arrange into tetratic 
or uniaxial nematic configurations, but orient randomly. These large clusters, however, easily form layered, smectic 
structures at high density. 

The fact that vortices seem to affect tetratic 
and smectic configurations differently is also interesting.
Tetratic configurations seem to be very stiff against the excitation of vortices \cite{Aranson}, since the two mutually orthogonal local directors
are inconsistent with a rotational velocity field. Therefore,
uniform tetratic states, which include point defects that restore the symmetry of the system, may exist for very long times in stable configurations.
On the other hand, smectic configurations are easily disrupted by vorticity and local shear modes, and uniform, well-defined steady states cannot
be formed because smectic order is `entropically' favoured, but competes with vortex excitations.

\acknowledgments

Financial support from MINECO (Spain) under grants FIS2013-47350-C5-1-R,
FIS2015-66523-P, MTM2012-39101 and MTM2015-63914-P, and ICMAT Severo Ochoa (Spain) under 
Contract No. SEV-2015-0554  are acknowledged.

\appendix

\section{Appendix}

The orientational and positional orders in the cell are quantified by the local and global uniaxial nematic, tetratic 
and smectic order parameters. The nematic order parameters can be obtained either through the orientational distribution
function, $h(\phi;{\bm R})$, or from the local nematic and tetratic 
order tensors, $Q({\bm R})$ and $T({\bm R})$, all calculated locally at the grid points ${\bm R}$
and averaging over circular regions of a radius conveniently set to $4L$.
$\phi$ is the angle between the long particle axis and a reference fixed axis defined as the $x$ axis, 
i.e. $\cos{\phi}=\hat{\bm e}\cdot\hat{\bm x}$. The distribution function and order tensors depend on the position ${\bm R}$ because 
particles will in general form domains with different orientational order in different regions of the cell and also
through the spatial dependence of the local director, $\hat{\bm n}({\bm R})\equiv(\cos\phi_0({\bm R}),\sin\phi_0({\bm R}))$.  
To find the local director, the local ordering tensor, with elements 
$\displaystyle Q_{\alpha\beta}({\bm R})=\left<2\hat{e}_{\alpha}\hat{e}_{\beta}-\delta_{\alpha\beta}\right>$, is diagonalised.
The local director points along the eigenvector associated with the highest eigenvalue, which can be identified as the uniaxial order 
parameter $q_2({\bm R})$. The local tetratic order parameter, $q_4({\bm R})$, can be obtained from the tetratic tensor, 
$T_{\alpha\beta\gamma\delta}({\bm R})=4\langle \hat{e}_{\alpha}\hat{e}_{\beta}\hat{e}_{\gamma}\hat{e}_{\delta}\rangle-\frac{1}{2}(\delta_{\alpha\beta}
\delta_{\gamma\delta}+\delta_{\alpha\gamma}\delta_{\beta\delta}+\delta_{\alpha\delta}\delta_{\beta\gamma})$.
Alternatively, the local nematic order parameters can be computed from $h(\phi;{\bm R})$ as
\begin{eqnarray}
q_k({\bm R}) &=& \left<\cos{k\phi}\right>_h\nonumber\\&=&
\int_0^{2\pi}d\phi h(\phi;{\bm R})\cos{k[\phi-\phi_0({\bm R})]}.
\label{qk}
\end{eqnarray}
In the experiments, results for these parameters obtained from the two routes are slightly different but qualitatively consistent.
To obtain a global distribution function for the whole cell, $h(\phi)$, the local distribution functions are superimposed
by rotating the local directors to a common reference axis. The global nematic and tetratic order parameters can be
obtained by averaging the local nematic $q_2({\bm R})$ and tetratic $q_4({\bm R})$ order parameters
over all the circular regions defined in the cell and over time,
\begin{eqnarray}
q_k=\left<q_k({\bm R})\right>_{{\bm R},t},\hspace{0.5cm}k=2,4,
\end{eqnarray}
Local and global smectic order parameters can be computed in the same way.
To avoid surface effects, a shell immediately next to the surface is excluded from all the calculations. The thickness of this shell
depends on the aspect ratio and experimental conditions (see Ref. \cite{daniel}).

Orientational order parameters can also be defined for the 
{\it clusters}. Once a cluster is identified, a long axis $\hat{\bm e}_c$ and a geometrical centre can be 
obtained. The former is defined simply by diagonalising the gyration tensor and taking the 
eigenvector associated to the highest eigenvalue.
If the maximum centre-to-centre interparticle distance is less than $L$, then $\hat{\bm e}_c=\bar{\hat{\bm e}}$; 
otherwise we take the perpendicular vector, $\hat{\bm e}_c=\bar{\hat{\bm e}}_{\perp}$. Using $\hat{\bm e}_c$ 
an orientational distribution function for the clusters, $h_c(\phi)$, and cluster order parameters, 
$q_k^{(c)}$, can be calculated.

To characterise smectic ordering, we have monitored the evolution of two parameters. One is the standard smectic order
parameter $\sigma_S({\bm R})$, defined locally as
\begin{eqnarray}
\sigma_S({\bm R})=\left|\left<\frac{1}{n}\sum_{i=1}^n e^{i{\bm Q}\cdot{\bm r}_i}\right>_T\right|.
\end{eqnarray}
Here $n$ is the number of cylinders inside the 
averaging region centred at ${\bm R}$,
${\bm r}_i$ is the position of one cylinder, and ${\bm Q}$ is a wavevector
(tuned to the wavelength of the smectic density wave as measured directly
on the images). $\left|\cdots\right|$ denote the modulus of the
complex argument. As usual, a global smectic order parameter can be defined
for the whole sample,
\begin{eqnarray}
\sigma_S=\left<\sigma_S({\bm R})\right>_{\bm R},
\end{eqnarray}
where the average is over all the regions defined in the cavity (excluding
a shell next to the circular wall) and over time.


\end{document}